%% file: MLMC.tex
\documentclass[11pt,a4paper]{article}
\usepackage{caption}
\usepackage[left=15mm, top=10mm, right=15mm, bottom=20mm, nohead=true, nofoot=true]{geometry}

\usepackage[utf8]{inputenc}
\usepackage{authblk}
\usepackage{indentfirst}
\usepackage{misccorr}
\usepackage{graphicx}
\usepackage{amsmath}
\usepackage{amssymb}
\usepackage{multicol}
\usepackage{bm}
\usepackage{longtable}
\usepackage{pdflscape}
\usepackage{epstopdf}
\usepackage{multirow}
\usepackage{adjustbox}
\usepackage{amsfonts}
\usepackage{ifthen}
\usepackage{fancyhdr}
\usepackage{setspace}
\usepackage{xcolor}
\usepackage[round,authoryear,comma,sort&compress,numbers]{natbib}
\usepackage{subcaption} 
\usepackage[toc,title,page]{appendix}
\usepackage[hidelinks]{hyperref}
\usepackage{url}

\hypersetup{
    colorlinks=true,
    linkcolor=green,
    citecolor=blue,
    filecolor=magenta,
    urlcolor=cyan,
    pdftitle={Sharelatex Example},
    pdfpagemode=FullScreen,
}

\fancypagestyle{plain}{\chead{\texttt{\textcolor{brown}{Communications of BAO, Vol.~71, Issue~2, 2024, pp.~377--382}}}}
\title{\textbf{Machine Learning Classification of Young Stellar Objects and Evolved Stars in the Magellanic Clouds Using the Probabilistic Random Forest Classifier}}
\author[1]{Sepideh Ghaziasgar \thanks{sepide.ghaziasgar@gmail.com, Corresponding author}}
\author[1]{Mahdi Abdollahi 
\thanks{m.abdollahi@ipm.ir}}
\author[1]{Atefeh Javadi 
\thanks{atefeh@ipm.ir}}
\author[2]{Jacco Th. van Loon \thanks{j.t.van.loon@keele.ac.uk}}
\author[3]{Iain McDonald \thanks{Iain.Mcdonald-2@manchester.ac.uk}}
\author[2]{Joana Oliveira \thanks{j.oliveira@keele.ac.uk}}
\author[1,5]{Habib G. Khosroshahi \thanks{habib@ipm.ir}}
\affil[1]{\scriptsize School of Astronomy, Institute for Research in Fundamental Sciences (IPM), P.O. Box 19568-36613, Tehran, Iran}
\affil[2]{\scriptsize Lennard-Jones Laboratories, Keele University, ST5 5BG, UK}
\affil[3]{\scriptsize Jodrell Bank Centre for Astrophysics, Alan Turing Building, University of Manchester, M13 9PL, UK}
\affil[5]{\scriptsize Iranian National Observatory, Institute for Research in Fundamental Sciences (IPM), Tehran, Iran}

\input jr.tex

\begin{document}
\pagestyle{empty}
\newpage
\pagestyle{fancy}
\label{firstpage}
\date{}
\maketitle

\begin{abstract}
The Magellanic Clouds (MCs) are excellent locations to study stellar dust emission and its contribution to galaxy evolution. Through spectral and photometric classification, MCs can serve as a unique environment for studying stellar evolution and galaxies enriched by dusty stellar point sources. We applied machine learning classifiers to spectroscopically labeled data from the Surveying the Agents of Galaxy Evolution (SAGE) project, which involved 12 multiwavelength filters and 618 stellar objects at the MCs. We classified stars into five categories: young stellar objects (YSOs), carbon-rich asymptotic giant branch (CAGB) stars, oxygen-rich AGB (OAGB) stars, red supergiants (RSG), and post-AGB (PAGB) stars. Following this, we augmented the distribution of imbalanced classes using the Synthetic Minority Oversampling Technique (SMOTE). Therefore, the Probabilistic Random Forest (PRF) classifier achieved the highest overall accuracy, reaching ${89\%}$ based on the recall metric, in categorizing dusty stellar sources before and after data augmentation. In this study, SMOTE did not impact the classification accuracy for the CAGB, PAGB, and RSG categories but led to changes in the performance of the OAGB and YSO classes.\\

\end{abstract}

\section{Introduction}

The Magellanic Clouds (MCs), as nearby dwarf galaxies with the distance of 50 kpc and 60 kpc \citep{Pietrzy-lmc-2013Natur.495...76P, LMC-SMC-2009A&A...496..399S,Subramanian-2011ASInC...3..144S} and metallicities of 0.5 and 0.2$Z_{\odot}$ \citep{1992ApJ-metalicity-russel}, offer an ideal environment to study stellar contributions to dust production \citep{2015MNRAS.451.3504R}.\\ 
\indent The life cycle of stars is represented by different stellar classes, each with distinct physical characteristics and processes. Dusty stellar objects enriched chemically during evolution can be classified into young stellar objects (YSOs) and evolved stars  \citep{Boyer2011b}.\\
\indent Young stellar objects (YSOs) are in the early phases of star formation. They are surrounded by gas and dust and offer a window into the physical processes driving star formation and galaxy evolution \citep{Sewilo-2013ApJ...778...15S,2023ApJ-yso-starformation,2016JASS...33..119S}. Also, the luminosity of YSOs can vary from optical to IR depending on their mass and evolution stage \citep{2018Ap&SS-Miettinen-ml-yso, oliveira-2013MNRAS.428.3001O}.\\
\indent Evolved stars, including asymptotic giant branch (AGB) stars with low– and intermediate mass (0.8-8 \(\textup{M}_\odot\)) and red supergiants (RSGs) with high mass (M$\ge$ 8 \(\textup{M}_\odot\)), are dust producers that enrich the ISM with heavy elements  \citep{2005ARA&A-falk-ahb, 2018A&AR-massloss-agb-hofner}. The significant brightness of AGBs $(\sim 10^{3-4}\  \textup{L}_\odot)$, along with their radial pulsations, makes these stars detectable in infrared \citep{Goldman-2017,2016ApJ-McDonald}. Many evolved AGBs are long-period variables (LPVs) \citep{Marigo-2017ApJ...835...77M,Torki-2023IAUS..362..353T, 2021ApJ-saremi2, Whitelock-2003MNRAS.342...86W, Navabi-2021ApJ...910..127N,2013-Javdi,Javadi-2019IAUS..343..283J,Javadi-2022IAUS}. AGB stars are classified into oxygen-rich (OAGB) and carbon-rich (CAGB) subcategories based on their chemical abundance, while RSGs represent massive stars in the final stages of their lives \citep{yang2020-smc,2010ASPC-Levesque-redsupergiant,2011-javadi,2003AJ-massey-redsuperginat}, often culminating in supernovae or compact remnants. Post-AGB (PAGB) stars mark a transitional phase, shedding their outer layers before evolving into white dwarfs, revealing unique chemical signatures \citep{2003ARA-vanWincke, 2020JApA-kmath-agb-postagb,Kamath2014MNRAS, Kamath2015MNRAS}.\\
\indent Machine learning algorithms are powerful tools for classifying stellar objects \citep{Ghaziasgar-2024eas..conf..371G,2022arXiv-ghaziasgar,2019arXiv190407248B, 2021MNRAS-Jacco2021-6822,2022MNRAS.517..140K}. The availability of more spectroscopically and photometrically labeled data enhances the ability of these algorithms to classify dusty stellar classes with greater accuracy, improving the overall reliability of stellar classification.\\ 
\section{Data}

The dataset used in this study is derived from the Surveying the Agents of Galaxy Evolution (SAGE) project for tracking dust and gas in Magellanic Clouds \citep{2006AJ-Meixner}. This dataset comprises multiwavelength spectroscopically labeled near-infrared and mid-infrared filters. From the SAGE spectral catalog, as shown in  Table~\ref{table:1}, we selected 12 multiwavelength filters for each spectral class (SpClass) and 618 dusty stellar objects in the MCs \citep{2011MNRAS.411.1597W, 2010PASP..122..683K,2015MNRAS.451.3504R, 2017MNRAS.470.3250J}. The features selected for training include UMmag, BMmag, VMmag, IMmag, J2mag, H2mag, Ks2mag, IRAC1, IRAC2, IRAC3, IRAC4, and [24]. \\
\indent We augmented our dataset using the SMOTE (Synthetic Minority Oversampling Technique) approach \citep{2011arXiv1106.1813C} that addresses the class imbalance in datasets as presented in Table~\ref{table:1}. Instead of simply duplicating instances from the minority class, SMOTE generates synthetic samples by interpolating between existing data points. This method selects a random data point from the minority class and creates new samples along the line segments connecting the sample to the nearest neighbor \citep{2011arXiv1106.1813C}. The SMOTE algorithm, applied to the training datasets, balances the population of minority classes with the majority class, potentially improving classifier performance. The Simple approach represents the original imbalanced dataset, while the SMOTE approach refers to the augmented dataset, where class imbalance has been addressed using SMOTE. \\
\begin{table}[h!]
\centering
     \caption{This is a spectral classification of dusty stars in the Magellanic Clouds. Based on SMOTE methodology, the ``Augmented Data'' column represents the population after data augmentation.}
    \label{table:1} 
    \begin{tabular}{ l c c c r }
        \hline 
        Classes & LMC & SMC & Total & *Augmented\\
            &   &   &   &Data\\
        \hline  
        CAGB & 136 & 38 & 174 & 200\\
        OAGB & 88 & 19 & 107 & 193 \\
        PAGB & 33 & 4 & 37 & 183\\
        RSG & 72 & 22 & 94 & 190\\
        YSO & 157 & 49 & 206 & 206\\
        \hline
    \end{tabular}
\end{table}
%
\section{Classification Models}
We employed supervised learning algorithms to classify samples of YSOs and evolved stars \citep{2022arXiv-ghaziasgar, Ghaziasgar-2024eas..conf..371G}. The algorithms were trained on spectroscopically labeled data and evaluated on a test dataset to assess their accuracy. The models we used included Probabilistic Random Forest (PRF) \citep{2019AJ....157...16R,2019arXiv190407248B,2021MNRAS-Jacco2021-6822, 2022MNRAS.517..140K}, Random Forest (RF) \citep{2001MachL..45....5B, 2010ApJ-randomforest-redshift,2017MNRAS-rf-dalya,2019arXiv190407248B}, K-Nearest Neighbor (KNN) \citep{knn1992}, C-Support Vector Classification (SVC) including SVC-poly and SVC-rbf \citep{vapnik95,2019arXiv190407248B}, and Gaussian Naive Bayes (GNB) \citep{2023MNRAS-yso-ml-naive-bayse}. Among all the classifiers, the PRF model performed best before and after data augmentation with the SMOTE method, as detailed below.\\
\indent The PRF classifier, the developed version of RF, is designed to handle noisy and uncertain datasets \citep{2019AJ....157...16R}. The RF is a machine learning algorithm that builds an ensemble of decision trees, each trained on a randomly selected subset of features and data, to avoid overfitting and generalize well to new data. In RF, predictions are made through majority voting for classification studies \citep{2001MachL..45....5B, 2019arXiv190407248B}. However, RF assumes that data and labels are fixed and accurate, making it less effective when dealing with noisy or uncertain inputs. The PRF overcomes this limitation by introducing a probabilistic framework that treats both features and labels as probability distributions rather than fixed values. PRF routes data points probabilistically across tree branches, accounting for uncertainties in both features and labels. This probabilistic framework enhances its ability to handle noisy inputs effectively \citep{2021MNRAS-Jacco2021-6822, 2022MNRAS.517..140K, 2019AJ....157...16R}.\\
\section{Results and Ongoing works}

\begin{figure}[h!]
    \centering
    \includegraphics[width=0.7\linewidth, clip]{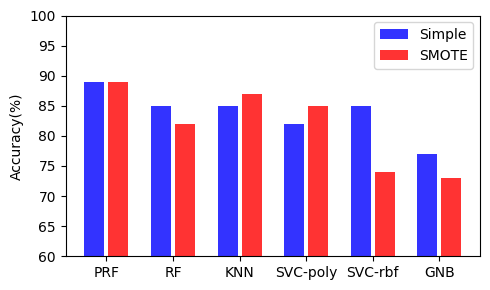}
    \caption{This plot presents the performance of Simple classifiers and  SMOTE classifiers based on their accuracy scores.}
    \label{fig:hist_model_results}
\end{figure}

\begin{figure*}[h!] 
    \centering
    \includegraphics[width=0.45\linewidth,height=0.45\linewidth]{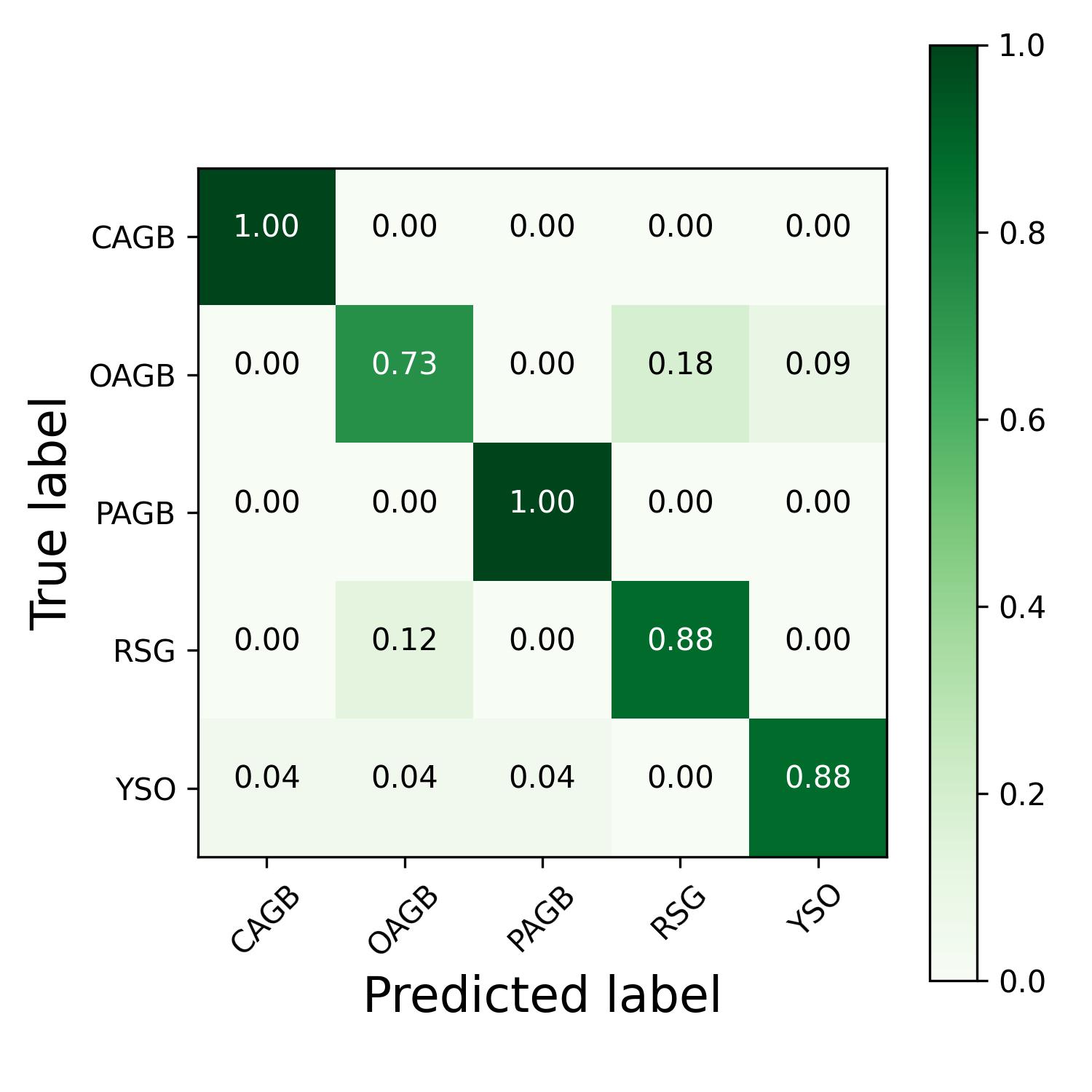}
    \hspace{0.05\linewidth} 
    \includegraphics[width=0.45\linewidth,height=0.45\linewidth]{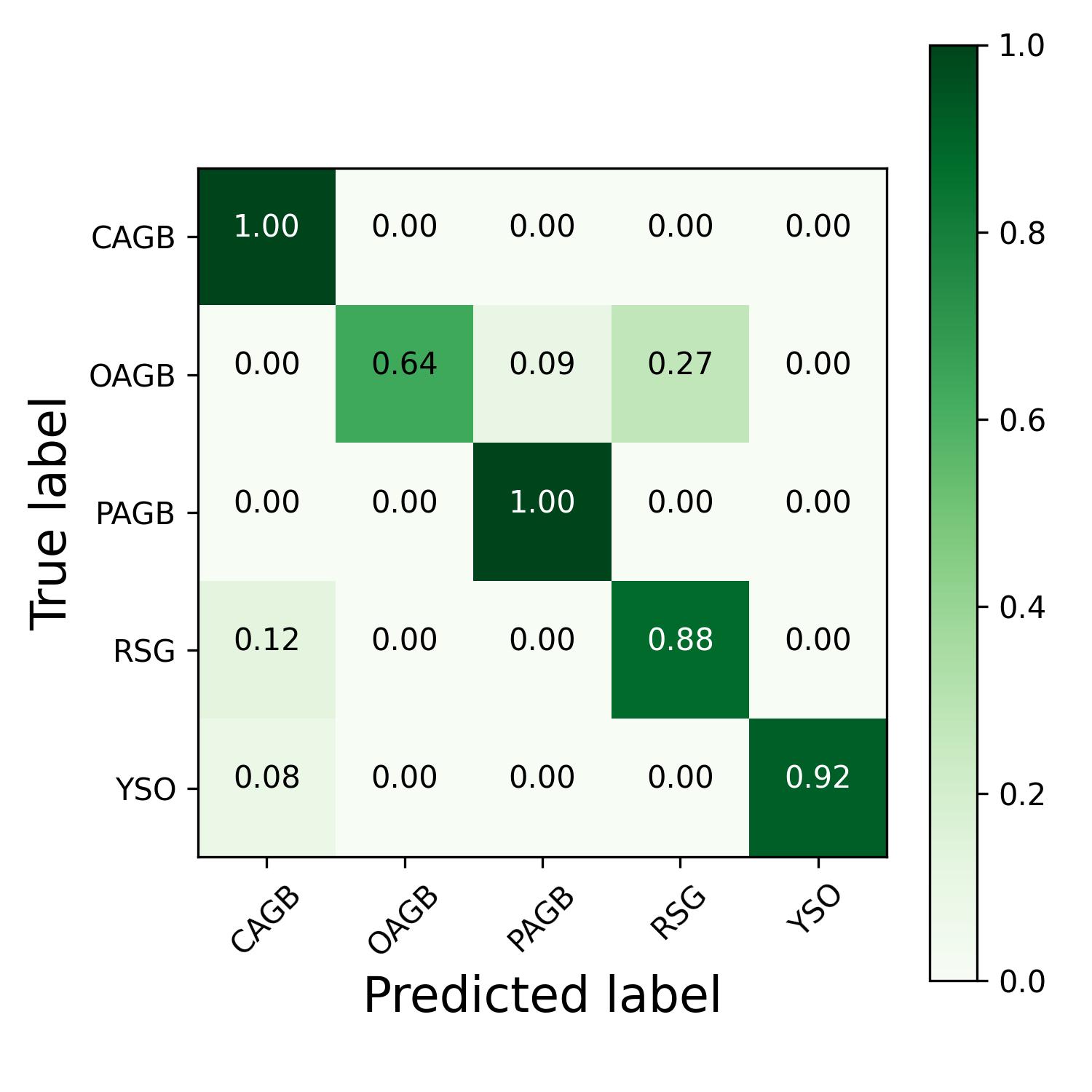}
    \caption{These are the confusion matrices for the Probabilistic Random Forest. The left panel presents the results before data augmentation, and the right panel displays the results after data augmentation with SMOTE. The matrix displays the number of objects predicted by the model in each class. The diagonal elements represent the predicted and actual labels for each class.}
    \label{fig:sub-PRF}
\end{figure*}

\indent We presented the classification results, as can be seen in Table~\ref{tab:overall_caption}, Fig.~\ref{fig:hist_model_results} and Fig.~\ref{fig:sub-PRF}, using two approaches, Simple and SMOTE, named based on the distribution of each dataset. As shown, in comparison to other classifiers, SMOTE outperforms Simple in PRF and RF classification. Based on the recall metric, the PRF classifier demonstrated the highest total accuracy, achieving ${89\%}$.
Using the SMOTE technique, the performance of the best model for the CAGB, PAGB, and RSG classes did not improve, with accuracy remaining at ${100\%}$, ${100\%}$ and ${88\%}$, respectively. However, SMOTE led to some variations in the OAGB and YSO classes.\\
\indent In the following, we can compare photometrically labeled data with spectroscopically labeled data with similar features. Additionally, incorporating multiwavelength data as model inputs could refine label determination for each object. More multiwavelength and spectroscopic observations are needed to improve dusty stellar classifications, especially for less populated classes like PAGBs and RSGs.\\
\begin{table}[h!]
    \centering
    \caption{Classification report; contains the model's precision, recall, and f1-score values for each class. The f1-score is calculated by averaging. The macro average f1-score represents an average of the f1-score over classes. The weighted average f1-score is calculated as the mean of all per-class f1-scores while considering each class’s support.}
    \label{tab:overall_caption}
    \begin{subtable}[t]{0.45\textwidth}
        \captionsetup{labelsep=period}
        \centering
        \caption{Classification report, Simple PRF.}
        \label{tab:PRF_simple}
        \begin{tabular}{l c c r}
            \hline
            Class & Precision & Recall & F1-score \\
            \hline
            CAGB & 0.95 & 1.00 & 0.97\\
            OAGB & 0.80 & 0.73 & 0.76\\  
            PAGB & 0.50 & 1.00 & 0.67\\
            RSG & 0.78 & 0.88 & 0.82\\
            YSO & 0.95 & 0.88 & 0.91\\
            \hline
            accuracy & & & 0.89 \\
            macro avg & 0.80 & 0.90 & 0.83 \\
            weighted avg & 0.89 & 0.89 & 0.89 \\
            \hline
        \end{tabular}
    \end{subtable}
    \hfill
    \begin{subtable}[t]{0.45\textwidth}
        \captionsetup{labelsep=period}
        \centering
        \caption{Classification report, SMOTE PRF.}
        \label{tab:PRF_smote}
        \begin{tabular}{l c c r}
            \hline
            Class & Precision & Recall & F1-score \\
            \hline
            CAGB & 0.86 & 1.00 & 0.92\\
            OAGB & 1.00 & 0.64 & 0.78\\  
            PAGB & 0.50 & 1.00 & 0.67\\
            RSG & 0.70 & 0.88 & 0.78\\
            YSO & 1.00 & 0.92 & 0.96\\
            \hline
            accuracy & & & 0.89 \\
            macro avg & 0.81 & 0.89 & 0.82 \\
            weighted avg & 0.91 & 0.89 & 0.89 \\
            \hline
        \end{tabular}
    \end{subtable}
\end{table}
\clearpage 

\section*{\small Acknowledgements}
\scriptsize{The authors thank the School of Astronomy at the Institute for Research in Fundamental Sciences (IPM) and the Iranian National Observatory (INO) for supporting this
research. S. Ghaziasgar is grateful for the support of the Byurakan Astrophysical Observatory (BAO).}

\scriptsize
\bibliographystyle{ComBAO}
\nocite{*}
\bibliography{MLMC}

\newpage
\appendix
\renewcommand{\thesection}{\Alph{section}.\arabic{section}}
\setcounter{section}{0}
\normalsize

\end{document}

%% file: jr.tex
\def\apj{Astrophys.~J. }

\def\pasp{Publ.~Astron.~Soc.~Pac. }

\def\aap{Astron.~Astrophys. }

\def\nat{Nature. }
\def\apjl{Astrophys.~J.~Lett. }

\def\aj{Astron.~J. }
\def\mnras{Mon.~Not.~R.~Astron.~Soc. }
\def\araa{Ann.~Rev.~Astron.~Astrophys. }

\def\apss{Astrophys.~Space.~Sci. }

\def\aapr{Astron.~Astrophys.~Rev. }

